\documentclass[showpacs,preprintnumbers,amsmath,amssymb]{revtex4}

\usepackage{graphicx}
\usepackage{bm}
\usepackage{color}

\newcommand{\bq}{\begin{equation}}
\newcommand{\eq}{\end{equation}}
\newcommand{\bqa}{\begin{eqnarray}}
\newcommand{\eqa}{\end{eqnarray}}
\newcommand{\nn}{\nonumber \\}

\def\be     {\begin{equation}}
\def\ee     {\end{equation}}
\def\bea        {\begin{eqnarray}}
\def\eea        {\end{eqnarray}}
\def\bnn    {\begin{eqnarray*}}
\def\enn    {\end{eqnarray*}}

\begin{document}

\title{Geometric encoding of renormalization group $\beta-$functions in an emergent holographic dual description}
\author{Ki-Seok Kim}
\affiliation{Department of Physics, POSTECH, Pohang, Gyeongbuk 37673, Korea}

\date{\today}

\begin{abstract}
We show that the information of renormalization group (RG) equations are geometrically encoded into an emergent holographic dual description, which duality-transforms a $D-$dimensional quantum field theory into a $(D+1)-$dimensional classical gravity theory in the large $N$ limit, where $N$ represents the flavor degeneracy of quantum fields. First, we reformulate RG transformations in a recursive way with introduction of an order-parameter field. As a result, we manifest the RG flow of an effective field theory through the emergence of an extra dimensional space, where both RG $\beta-$functions of coupling functions and RG flow equation of the order-parameter field appear in the resulting effective action explicitly through the extra dimensional space. Second, we consider an effective dual holographic description derived recently, where the classical gravity theory of the large $N$ limit takes into account quantum corrections in the all-loop order. This non-perturbative nature turns out to originate from an intertwined renormalization structure between both RG flow equations of coupling functions and order-parameter fields in the emergent extra-dimensional space, where the IR boundary condition of the order-parameter field gives rise to a mean-field equation with fully renormalized interaction coefficients. Third, comparing the RG-reformulated effective field theory with this effective dual holographic description, we obtain term-by-term matching conditions in the level of an effective action. As a result, we express all RG coefficients of fields and interaction vertices in terms of the metric tensor of the dual holographic theory. Through this metric reformulation for the RG analysis, we propose a prescription on how to find RG $\beta-$functions of interaction coefficients in a non-perturbative way beyond the perturbative RG analysis. In particular, we claim that the present prescription of the RG flow generalizes the holographic RG flow of the holographic duality conjecture towards the absence of conformal symmetry, where the emergent holographic dual effective field theory has been derived from the first principle, thus being applicable even away from quantum criticality.
\end{abstract}


\maketitle

\section{Introduction}

Holographic duality conjecture \cite{Holographic_Duality_I,Holographic_Duality_II,Holographic_Duality_III,Holographic_Duality_IV,Holographic_Duality_V, Holographic_Duality_VI,Holographic_Duality_VII} states that a $D-$dimensional quantum field theory can be duality-mapped into a $(D+1)-$dimensional quantum gravity theory, where the quantum gravity theory is a hologram of the quantum field theory. Correlation functions of conserved currents in the quantum field theory are given by those of holographically dual dynamical fields in the quantum gravity theory, where energy-momentum tensor and U(1) charge currents are dual holographically described by metric tensor and U(1) gauge fields, respectively. The real power of this conjecture is based on the fact that quantum fluctuations of such bulk dual fields as metric tensor and U(1) gauge fields are suppressed to become classical in the so called large$-N$ limit, where $N$ corresponds to the number of degrees of freedom, central charge, roughly speaking. As a result, the dynamics of strongly coupled quantum field theories is described by a holographically realized classical gravity theory with bulk gauge fields.

It turns out that this dual holographic description predicts a novel quantum liquid state, referred to as holographic liquid \cite{Holographic_Liquid_Son_I,Holographic_Liquid_Son_II,Holographic_Liquid_Son_III, Holographic_Liquid_Son_IV}. Based on a vacuum solution for the metric tensor and U(1) gauge field, one may consider small variations in both metric tensor and U(1) gauge fields and find their normal mode solutions, which contain the dynamics information of energy-momentum tensor and U(1) charge currents. These classical solutions describe not only diffusive dynamics in transverse modes but also sound dynamics in longitudinal modes, which correspond to those modes in hydrodynamics \cite{Holographic_Liquid_Son_I,Holographic_Liquid_Son_II,Holographic_Liquid_Son_III}. In particular, this dual holographic description gives rise to a universally small value of the $\eta / s$ ratio, where $\eta$ is shear viscosity and $s$ is entropy \cite{Holographic_Liquid_Son_IV}. Furthermore, one may extract out the Lyapunov exponent \cite{Lyapunov_Exponent} from out-of-time-ordered correlation functions \cite{OTOC} of dual holographic fields, known to be a measure of chaos. It has been conjectured that the Lyapunov exponent shows its maximum value in the holographic liquid state, referred to as chaos bound \cite{Chaos_Bound_I,Chaos_Bound_II,Chaos_Bound_III,Chaos_Bound_IV}.

It may not be so surprising that the dual holographic description gives rise to effective hydrodynamics, where only energy-momentum tensor and U(1) charge currents are slow variables to dominate over the low-energy physics. However, neither the universally small value of $\eta/s$ nor the chaos bound of the Lyapunov exponent is within the perturbative theoretical framework as far as we understand. We believe that the dynamic nature of the holographic liquid is beyond the symmetry principle. The question that we try to address in this paper is how solving classical equations of motion coupled to background gravity with an extra dimensional space gives rise to non-perturbative physics in strongly coupled quantum field theories.

To shed light on this issue, it is necessary to construct an effective dual holographic theory from a quantum field theory based on the first principle \cite{RG_Holography_I,RG_Holography_II,RG_Holography_III,RG_Holography_IV,RG_Holography_V,RG_Holography_VI,RG_Holography_VII,RG_Holography_VIII,
RG_Holography_IX,RG_Holography_X,RG_Holography_XI,RG_Holography_XII,RG_Holography_XIII,RG_Holography_XIV,RG_Holography_XV,RG_Holography_XVI,
RG_Holography_XVII,RG_Holography_XVIII,RG_Holography_XIX,RG_Holography_XX,RG_Holography_XXI,SungSik_Holography_I,SungSik_Holography_II,
SungSik_Holography_III,Holographic_Description_Entanglement_Entropy,Holographic_Description_Kondo_Effect,Holographic_Description_Einstein_Klein_Gordon,
Holographic_Description_Einstein_Maxwell}. An idea is to introduce a renormalization group (RG) energy scale into the effective field theory as an extra dimensional space and to manifest the RG flow of the effective field theory through the extra dimensional space. Actually, this idea has been investigated in the holographic duality conjecture, referred to as holographic renormalization \cite{Holographic_Duality_IV,Holographic_Duality_V, Holographic_Duality_VI}. RG flows of interaction vertices are realized in the extra dimensional space, serving as an effective curved spacetime for renormalized low-energy fluctuations in this effective field theory. RG flows of correlation functions can be described by introducing order parameter fields into the effective field theory, giving rise to a holographic dual formulation in the presence of this effective curved spacetime background. See Fig. 1 of Ref. \cite{Holography_RG_Concept} for conceptual realization of this idea.

Recently, we proposed how to derive a holographic dual effective field theory from a quantum field theory: Einstein-Klein-Gordon-type effective field theory in terms of dynamical metric tensor and dual scalar fields from interacting scalar fields \cite{Holographic_Description_Einstein_Klein_Gordon} and Einstein-Maxwell-type effective field theory in terms of dynamical metric tensor and U(1) gauge fields from interacting Dirac fermions \cite{Holographic_Description_Einstein_Maxwell}, respectively. Following the idea of the holographic RG flow, we implemented Wilsonian RG transformations \cite{RG_Textbook} recursively not in momentum space but in real space a la Polchinski \cite{RG_realspace_Polchinski}, also developed further by Sung-Sik Lee in his emergent holographic construction \cite{SungSik_Holography_I,SungSik_Holography_II,SungSik_Holography_III}. This recursive RG transformation technique may be regarded as an inverse procedure of the holographic renormalization \cite{Holographic_Duality_IV,Holographic_Duality_V, Holographic_Duality_VI}. Quantum fluctuations of both metric tensor and holographic dual (scalar or gauge) fields are frozen to cause a classical field theory in the large $N$ limit, where $N$ is the flavor degeneracy of quantum matter fields. It turns out that the classical gravity theory of the large $N$ limit takes into account quantum corrections in the all-loop order \cite{Holographic_Description_Kondo_Effect}. This non-perturbative nature originates from an intertwined renormalization structure between both RG flow equations of coupling functions and order-parameter fields in the emergent extra-dimensional space, where the IR boundary condition of the order-parameter field gives rise to a mean-field equation with fully renormalized interaction coefficients \cite{Intertwined_RG_Structure_KSK}.

In the present study, we clarify how the information of RG equations are geometrically encoded into the emergent holographic dual description. More precisely, we express all RG coefficients of fields and interaction vertices in terms of metric tensor of the dual holographic theory. Through this metric reformulation for the RG analysis, we propose a prescription on how to find RG $\beta-$functions of interaction coefficients in a non-perturbative way beyond the perturbative RG analysis. In particular, we claim that the present prescription of the RG flow generalizes the holographic RG flow of the holographic duality conjecture towards the absence of conformal symmetry, where the emergent holographic dual effective field theory has been derived from the first principle, thus being applicable even away from quantum criticality \cite{Holographic_Description_Entanglement_Entropy_Comments}.

\section{Reformulation of renormalization group transformations in a recursive way for interacting relativistic bosons}

\subsection{Renormalization group analysis}

We start our discussions, reviewing the RG method \cite{RG_Textbook}. An effective field theory for interacting scalar fields is given by
\bqa && Z = \int D \phi_{\alpha}^{(B)}(x) \exp\Big[ - \int d^{D} x \Big\{ [\partial_{\mu} \phi_{\alpha}^{(B)}(x)]^{2} + m_{B}^{2} [\phi_{\alpha}^{(B)}(x)]^{2} + \frac{u_{B}}{2N} [\phi_{\alpha}^{(B)}(x)]^{2} [\phi_{\beta}^{(B)}(x)]^{2} \Big\} \Big] . \eqa
$\phi_{\alpha}^{(B)}(x)$ is a real scalar field with a flavor index $\alpha = 1, ..., N$ at a $D-$dimensional spacetime coordinate $x$, where the superscript $(B)$ denotes a bare field before renormalization. $m_{B}$ is the bare mass of these scalar bosons and $u_{B}$ is the strength of their self-interactions.

The RG transformation is a careful way of the path integral. First, we separate the original bare field into low-energy and high-energy degrees of freedom. This mode separation can be performed either in energy-momentum space or in real spacetime. For example, the high-energy mode is defined within a thin shell in the momentum space. Now, we take the path integral for the high-energy degrees of freedom, where their quantum fluctuations give rise to renormalization in interaction vertices between such low-energy degrees of freedom. Rescaling not only the spacetime coordinate but also both low-energy quantum fields and their renormalized interactions, which return the resulting effective action into its original form, we obtain an effective field theory in terms of renormalized fields with renormalized interactions
\bqa && Z = \int D \phi_{\alpha}^{(R)}(y) \exp\Big[ - \int d^{D} y \Big\{ Z_{\phi} [\partial_{\mu} \phi_{\alpha}^{(R)}(y)]^{2} + Z_{m^2} m_{R}^{2} [\phi_{\alpha}^{(R)}(y)]^{2} + Z_{u} \frac{u_{R}}{2N} [\phi_{\alpha}^{(R)}(y)]^{2} [\phi_{\beta}^{(R)}(y)]^{2} \Big\} \Big] . \label{Conventional_RG_Partition_Function} \eqa
The spacetime coordinate $x$ is rescaled to be $x = e^{- l} y$, where $l$ is an infinitesimal scaling parameter, identified with an RG energy scale. $\phi_{\alpha}^{(R)}(y)$ is a renormalized field, coming from the low-energy part of the bare field. Quantum fluctuations of the high-energy modes give rise to self-energy corrections in low-energy fluctuations, encoded into the field-renormalization constant $Z_{\phi}$. The RG equation for the quantum field is given by
\bqa && \phi_{\alpha}^{(B)}(x) = e^{\frac{D-2}{2} l} Z_{\phi}^{\frac{1}{2}} \phi_{\alpha}^{(R)}(y) . \eqa
The mass of low-energy quantum fields is also renormalized, where the static part of the self-energy correction is translated into the mass-renormalization constant $Z_{m^{2}}$. The RG equation for the mass is given by
\bqa && m_{B}^{2}(x) = e^{2 l} Z_{\phi}^{-1} Z_{m^2} m_{R}^{2}(y) . \eqa
Renormalization in the interaction vertex between renormalized quantum fields is described by the RG coefficient $Z_{u}$, where the corresponding RG equation is given by
\bqa && u_{B}(x) = e^{(4 - D) l} Z_{\phi}^{-2} Z_{u} u_{R}(y) . \eqa
%
%

It is straightforward to rewrite these RG equations in the form of differential equations, given by
\bqa && \frac{d \ln m_{R}^{2}}{d l} = 2 - \frac{d \ln Z_{\phi}}{d l} + \frac{d \ln Z_{m^{2}}}{d l} \equiv \beta_{m^{2}}(m_{R}^{2}, u_{R}) , \nn && \frac{d \ln u_{R}}{d l} = 4 - D - 2 \frac{d \ln Z_{\phi}}{d l} + \frac{d \ln Z_{u}}{d l} \equiv \beta_{u}(m_{R}^{2}, u_{R}) , \eqa
and called RG $\beta-$functions for the interaction vertices of the quantum field theory. These RG $\beta-$functions show how renormalized interaction parameters evolve as a function of the RG energy scale $l$ to control quantum fluctuations of high-energy modes. As clearly shown in these equations, essential ingredients are the RG coefficients such as $Z_{\phi}$, $Z_{m^{2}}$, and $Z_{u}$. These RG coefficients are calculated based on the perturbation method. In this study we propose a prescription on how to determine them non-perturbatively based on an emergent and derived holographic dual effective field theory.

\subsection{Recursive renormalization group transformations}

To express all the RG coefficients in terms of the metric tensor, it is necessary to reformulate the above RG procedure in a recursive way. We rewrite the partition function Eq. (\ref{Conventional_RG_Partition_Function}) in the following way
\bqa && Z = \int D \phi_{\alpha}^{(R)}(y) D m^{(0) 2}(y) D \gamma_{2}^{(0)}(y) D u^{(0)}(y) D \gamma_{4}^{(0)}(y) \nn && \exp\Big[ - \int d^{D} y \Big\{ Z_{\phi}^{(0)} [\partial_{\mu} \phi_{\alpha}^{(R)}(y)]^{2} + Z_{m^2}^{(0)} m^{(0) 2}(y) [\phi_{\alpha}^{(R)}(y)]^{2} + Z_{u}^{(0)} \frac{u^{(0)}(y)}{2N} [\phi_{\alpha}^{(R)}(y)]^{2} [\phi_{\beta}^{(R)}(y)]^{2} \nn && + \gamma_{2}^{(0)}(y) [m^{(0) 2}(y) - e^{- 2 l} Z_{\phi}^{(0)} Z_{m^2}^{(0) -1} m_{B}^{2}] + \gamma_{4}^{(0)}(y) [u^{(0)}(y) - e^{- (4 - D) l} Z_{\phi}^{(0) 2} Z_{u}^{(0) -1} u_{B}] \Big\} \Big] . \eqa
$\gamma_{2}^{(0)}(y)$ and $\gamma_{4}^{(0)}(y)$ are Lagrange multiplier fields to impose the RG-equation constraints for the mass and self-interaction parameter, respectively. The reason for the introduction of the superscript $(0)$ will be clarified soon.

%
%

Once again, we rewrite the above expression as follows
\bqa && Z = \int D \phi_{\alpha}^{(R)}(y) D m^{(0) 2}(y) D m^{(1) 2}(y) D \gamma_{2}^{(0)}(y) D \gamma_{2}^{(1)}(y) D u^{(0)}(y) D u^{(1)}(y) D \gamma_{4}^{(0)}(y) D \gamma_{4}^{(1)}(y) \nn && \exp\Big[ - \int d^{D} y Z_{\phi}^{(0)} \Big\{ [\partial_{\mu} \phi_{\alpha}^{(R)}(y)]^{2} + Z_{m^2}^{(0) -1} m^{(1) 2}(y) [\phi_{\alpha}^{(R)}(y)]^{2} + Z_{u}^{(0) -1} \frac{u^{(1)}(y)}{2N} [\phi_{\alpha}^{(R)}(y)]^{2} [\phi_{\beta}^{(R)}(y)]^{2} \Big\} \nn && - \int d^{D} y Z_{\phi}^{(0)} \Big\{ \gamma_{2}^{(0)}(y) [m^{(0) 2}(y) - e^{- 2 l} m_{B}^{2}] + \gamma_{4}^{(0)}(y) [u^{(0)}(y) - e^{- (4 - D) l} u_{B}] \Big\} \nn && - \int d^{D} y Z_{\phi}^{(0)} \Big\{ \gamma_{2}^{(1)}(y) [m^{(1) 2}(y) - Z_{m^2}^{(0)} m^{(0) 2}(y)] + \gamma_{4}^{(1)}(y) [u^{(1)}(y) - Z_{\phi}^{(0)} Z_{u}^{(0)} u^{(0)}(y)] \Big\} + 4 \int d^{D} y \ln Z_{\phi}^{(0)} \Big] , \label{First_Iteration_RG_without_varphi} \eqa
where $\gamma_{2}^{(1)}(y)$ and $\gamma_{4}^{(1)}(y)$ are Lagrange multiplier fields to impose the constraint equations. The last term results from the existence of $Z_{\phi}^{(0)}$ in the $\delta-$function constraints of RG equations. It is straightforward to check that this expression recovers Eq. (\ref{Conventional_RG_Partition_Function}) after the path integral of $\int D m^{(0) 2}(y) D m^{(1) 2}(y) D \gamma_{2}^{(0)}(y) D \gamma_{2}^{(1)}(y) D u^{(0)}(y) D u^{(1)}(y) D \gamma_{4}^{(0)}(y) D \gamma_{4}^{(1)}(y)$, where all the RG coefficients are identified as follows
\bqa && Z_{\phi} = Z_{\phi}^{(0)} , ~~~~~ Z_{m^{2}} = Z_{m^{2}}^{(0)} , ~~~~~ Z_{u} = Z_{u}^{(0)} . \eqa
In this respect there is no further information here beyond Eq. (\ref{Conventional_RG_Partition_Function}).

Based on this reformulation, we suggest the following expression of the partition function
\bqa && Z = \int D \phi_{\alpha}^{(R)}(y) \Pi_{k = 0}^{f} D m^{(k) 2}(y) D \gamma_{2}^{(k)}(y) D u^{(k)}(y) D \gamma_{4}^{(k)}(y) \exp\Big[ - \int d^{D} y \Pi_{k = 1}^{f} Z_{\phi}^{(k-1)} \Big\{ [\partial_{\mu} \phi_{\alpha}^{(R)}(y)]^{2} \nn && + \Pi_{k = 1}^{f} Z_{m^{2}}^{(k-1) - 1} m^{(f) 2}(y) [\phi_{\alpha}^{(R)}(y)]^{2} + \Pi_{k = 1}^{f} Z_{u}^{(k-1) - 1} \frac{u^{(f)}(y)}{2N} [\phi_{\alpha}^{(R)}(y)]^{2} [\phi_{\beta}^{(R)}(y)]^{2} \Big\} \nn && - \int d^{D} y Z_{\phi}^{(0)} \Big\{ \gamma_{2}^{(0)}(y) [m^{(0) 2}(y) - e^{- 2 l} m_{B}^{2}] + \gamma_{4}^{(0)}(y) [u^{(0)}(y) - e^{- (4 - D) l} u_{B}] \Big\} \nn && - \int d^{D} y \sum_{k = 1}^{f} Z_{\phi}^{(k-1)} \Big\{ \gamma_{2}^{(k)}(y) [m^{(k) 2}(y) - Z_{m^2}^{(k-1)} m^{(k-1) 2}(y)] + \gamma_{4}^{(k)}(y) [u^{(k)}(y) - Z_{\phi}^{(k-1)} Z_{u}^{(k-1)} u^{(k-1)}(y)] \Big\} \nn && + 2 \int d^{D} y \ln Z_{\phi}^{(0)} + 2 \int d^{D} y \sum_{k = 1}^{f} \ln Z_{\phi}^{(k-1)} \Big] . \eqa
Here, identifying all the RG coefficients of Eq. (\ref{Conventional_RG_Partition_Function}) in the following way
\bqa && Z_{\phi} = \Pi_{k = 1}^{f} Z_{\phi}^{(k-1)} , ~~~~~ Z_{m^{2}} = Z_{m^{2}}^{(f-1)} , ~~~~~ Z_{u} = Z_{u}^{(f-1)} , \eqa
we also reproduce the expression of Eq. (\ref{Conventional_RG_Partition_Function}) from the above partition function.

Taking the limit of $l \rightarrow 0$ in this expression, we obtain
\bqa && Z = \int D \phi_{\alpha}^{(R)}(y) \Pi_{k = 0}^{f} D m^{(k) 2}(y) D \gamma_{2}^{(k)}(y) D u^{(k)}(y) D \gamma_{4}^{(k)}(y) \exp\Big[ - \int d^{D} y \exp\Big( d l \sum_{k = 1}^{f} \frac{d \ln Z_{\phi}^{(k-1)}}{d l}\Big) \Big\{ [\partial_{\mu} \phi_{\alpha}^{(R)}(y)]^{2} \nn && + \exp\Big(- d l \sum_{k = 1}^{f} \frac{d \ln Z_{m^{2}}^{(k-1)}}{d l}\Big) m^{(f) 2}(y) [\phi_{\alpha}^{(R)}(y)]^{2} + \exp\Big(- d l \sum_{k = 1}^{f} \frac{d \ln Z_{u}^{(k-1)}}{d l}\Big) \frac{u^{(f)}(y)}{2N} [\phi_{\alpha}^{(R)}(y)]^{2} [\phi_{\beta}^{(R)}(y)]^{2} \Big\} \nn && - \int d^{D} y Z_{\phi}^{(0)} \Big\{ d l \gamma_{2}^{(0)}(y) \Big( \frac{m^{(0) 2}(y) - m_{B}^{2}}{d l} + 2 m_{B}^{2} \Big) + d l \gamma_{4}^{(0)}(y) \Big( \frac{u^{(0)}(y) - u_{B}}{d l} + (4 - D) u_{B} \Big) \Big\} \nn && - \int d^{D} y d l \sum_{k = 1}^{f} Z_{\phi}^{(k-1)} \Big\{ \gamma_{2}^{(k)}(y) \Big( \frac{m^{(k) 2}(y) - m^{(k-1) 2}(y)}{d l} - \frac{d \ln Z_{m^2}^{(k-1)}}{d l} m^{(k-1) 2}(y) \Big) \nn && + \gamma_{4}^{(k)}(y) \Big( \frac{u^{(k)}(y) - u^{(k-1)}(y)}{d l} - \Big( \frac{d \ln Z_{\phi}^{(k-1)}}{d l} + \frac{d \ln Z_{u}^{(k-1)}}{d l} \Big) u^{(k-1)}(y) \Big) \Big\} \nn && + 2 \int d^{D} y \ln Z_{\phi}^{(0)} + 2 \int d^{D} y d l \sum_{k = 1}^{f} \frac{d \ln Z_{\phi}^{(k-1)}}{d l} \Big] , \eqa
where all terms were kept up to the linear order in $d l$.

It is interesting to observe
\bqa && d l \sum_{k = 1}^{f} \Longrightarrow \int_{0}^{z_{f}} d z . \eqa
Based on this replacement, we reformulate the partition function of Eq. (\ref{Conventional_RG_Partition_Function}) as follows
\bqa && Z = \int D \phi_{\alpha}^{(R)}(y) D m^{2}(y,z) D \gamma_{2}(y,z) D u(y,z) D \gamma_{4}(y,z) \nn && \exp\Big[ - \int d^{D} y \frac{Z_{\phi}(y,z_{f})}{Z_{\phi}(y,0)} \Big\{ [\partial_{\mu} \phi_{\alpha}^{(R)}(y)]^{2} + \frac{Z_{m^{2}}^{-1}(y,z_{f})}{Z_{m^{2}}^{-1}(y,0)} m^{2}(y,z_{f}) [\phi_{\alpha}^{(R)}(y)]^{2} + \frac{Z_{u}^{-1}(y,z_{f})}{Z_{u}^{-1}(y,0)} \frac{u(y,z_{f})}{2N} [\phi_{\alpha}^{(R)}(y)]^{2} [\phi_{\beta}^{(R)}(y)]^{2} \Big\} \nn && - \int d^{D} y Z_{\phi}(y,0) \Big\{ \gamma_{2}(y,0) \Big( \frac{\partial m^{2}(y,z)}{\partial z} \Big|_{z = 0} - 2 m^{2}(y,0) \Big) + \gamma_{4}(y,0) \Big( \frac{\partial u(y,z)}{\partial z} \Big|_{z = 0} - (4 - D) u(y,0) \Big) \Big\} \nn && - \int_{0}^{z_{f}} d z \int d^{D} y Z_{\phi}(y,z) \Big\{ \gamma_{2}(y,z) \Big( \frac{\partial m^{2}(y,z)}{\partial z} - \frac{\partial \ln Z_{m^2}(y,z)}{\partial z} m^{2}(y,z) \Big) \nn && + \gamma_{4}(y,z) \Big( \frac{\partial u(y,z)}{\partial z} - \Big( \frac{\partial \ln Z_{\phi}(y,z)}{\partial z} + \frac{\partial \ln Z_{u}(y,z)}{\partial z} \Big) u(y,z) \Big) \Big\} + 2 \int d^{D} y \ln Z_{\phi}(y,z_{f}) \Big] , \eqa
where the identification of RG coefficients is given by
\bqa && Z_{\phi} = Z_{\phi}(y,z_{f}) , ~~~~~ Z_{m^{2}} = Z_{m^{2}}(y,z_{f}) , ~~~~~ Z_{u} = Z_{u}(y,z_{f}) \eqa
when all initial values of RG coefficients are set to be $Z_{\phi}(y,0) = Z_{m^{2}}(y,0) = Z_{u}(y,0) = 1$. Now, the RG scale appears as an extra dimensional space in the level of the effective action. As a result, the RG flow of the effective action is manifested through the introduction of the extra dimensional space, where renormalized interaction parameters serve as an effective curved spacetime for the renormalized scalar fields. However, we emphasize that this partition function does not have any more information than the original expression in terms of renormalized fields and interaction vertices.

\subsection{Introduction of an order parameter field in the recursive renormalization group transformations}

To introduce non-perturbative physics into the previous RG procedure, we modify the RG transformation to take into account an intertwined renormalization structure between RG $\beta-$functions of interaction parameters and RG flow equation of an order-parameter field \cite{Intertwined_RG_Structure_KSK}. Although it is not clear what the intertwined renormalization structure means at present, we clarify this point in the discussion of an emergent and derived holographic dual effective field theory of the next section.

We introduce an order parameter field into Eq. (\ref{First_Iteration_RG_without_varphi}), taking the Hubbard-Stratonovich transformation for the self-interaction channel. Then, we obtain
\bqa && Z = \int D \phi_{\alpha}^{(R)}(y) D \varphi^{(0)}(y) D m^{(0) 2}(y) D m^{(1) 2}(y) D \gamma_{2}^{(0)}(y) D \gamma_{2}^{(1)}(y) D u^{(0)}(y) D u^{(1)}(y) D \gamma_{4}^{(0)}(y) D \gamma_{4}^{(1)}(y) \nn && \exp\Big[ - \int d^{D} y Z_{\phi}^{(0)} \Big\{ [\partial_{\mu} \phi_{\alpha}^{(R)}(y)]^{2} + Z_{m^{2}}^{(0) - 1} m^{(1) 2}(y) [\phi_{\alpha}^{(R)}(y)]^{2} - i \varphi^{(0)}(y) [\phi_{\alpha}^{(R)}(y)]^{2} + \frac{N Z_{u}^{(0)}}{2 u^{(1)}(u)} [\varphi^{(0)}(y)]^{2} \Big\} \nn && - \int d^{D} y Z_{\phi}^{(0)} \Big\{ \gamma_{2}^{(0)}(y) [m^{(0) 2}(y) - e^{- 2 l} m_{B}^{2}] + \gamma_{4}^{(0)}(y) [u^{(0)}(y) - e^{- (4 - D) l} u_{B}] \Big\} \nn && - \int d^{D} y Z_{\phi}^{(0)} \Big\{ \gamma_{2}^{(1)}(y) [m^{(1) 2}(y) - Z_{m^2}^{(0)} m^{(0) 2}(y)] + \gamma_{4}^{(1)}(y) [u^{(1)}(y) - Z_{\phi}^{(0)} Z_{u}^{(0)} u^{(0)}(y)] \Big\} + 4 \int d^{D} y \ln Z_{\phi}^{(0)} \Big] , \eqa
where $\varphi^{(0)}(y)$ is an order parameter field.

An important additional step is to perform the RG transformation for this newly introduced collective field. Separating the dual scalar field into low and high energy modes, and integrating over high-energy fluctuations, we obtain newly generated effective self-interactions of original scalar fields. Performing the Hubbard-Stratonovich transformation for this additional interaction channel once again, we obtain an independent new order parameter, denoted by $\varphi^{(1)}(y)$ in addition to $\varphi^{(0)}(y)$. As a result, we have the following expression
\bqa && Z = \int D \phi_{\alpha}^{(R)}(y) D \varphi^{(0)}(y) D \varphi^{(1)}(y) D m^{(0) 2}(y) D m^{(1) 2}(y) D \gamma_{2}^{(0)}(y) D \gamma_{2}^{(1)}(y) D u^{(0)}(y) D u^{(1)}(y) D \gamma_{4}^{(0)}(y) D \gamma_{4}^{(1)}(y) \nn && \exp\Big[ - \int d^{D} y Z_{\phi}^{(0)} \Big\{ [\partial_{\mu} \phi_{\alpha}^{(R)}(y)]^{2} + Z_{m^{2}}^{(0) - 1} m^{(1) 2}(y) [\phi_{\alpha}^{(R)}(y)]^{2} - i e^{- l} \varphi^{(1)}(y) [\phi_{\alpha}^{(R)}(y)]^{2} \nn && + \frac{N Z_{u}^{(0)}}{2 u^{(1)}(y)} [\varphi^{(0)}(y)]^{2} + l \frac{N Z_{u}^{(0)}}{2 u^{(1)}(y)} \Big( \frac{\varphi^{(1)}(y) - \varphi^{(0)}(y)}{l} \Big)^{2} \Big\} \nn && - \int d^{D} y Z_{\phi}^{(0)} \Big\{ \gamma_{2}^{(0)}(y) [m^{(0) 2}(y) - e^{- 2 l} m_{B}^{2}] + \gamma_{4}^{(0)}(y) [u^{(0)}(y) - e^{- (4 - D) l} u_{B}] \Big\} \nn && - \int d^{D} y Z_{\phi}^{(0)} \Big\{ \gamma_{2}^{(1)}(y) [m^{(1) 2}(y) - Z_{m^2}^{(0)} m^{(0) 2}(y)] + \gamma_{4}^{(1)}(y) [u^{(1)}(y) - Z_{\phi}^{(0)} Z_{u}^{(0)} u^{(0)}(y)] \Big\} + 4 \int d^{D} y \ln Z_{\phi}^{(0)} \Big] , \eqa
where the field shift of $\varphi^{(1)}(y) \Longrightarrow \varphi^{(1)}(y) - \varphi^{(0)}(y)$ has been taken. We note that $\varphi^{(0)}(y)$ is updated as $\varphi^{(1)}(y)$ in the coupling term of $- i e^{- l} \varphi^{(1)}(y) [\phi_{\alpha}^{(R)}(y)]^{2}$.

Following the previous RG strategy, we repeat this RG transformation and obtain the partition function
\bqa && Z = \int D \phi_{\alpha}^{(R)}(y) \Pi_{k = 0}^{f} D \varphi^{(k)}(y) D m^{(k) 2}(y) D \gamma_{2}^{(k)}(y) D u^{(k)}(y) D \gamma_{4}^{(k)}(y) \nn && \exp\Big[ - \int d^{D} y \Pi_{k = 1}^{f} Z_{\phi}^{(k-1)} \Big\{ [\partial_{\mu} \phi_{\alpha}^{(R)}(y)]^{2} + \Pi_{k = 1}^{f} Z_{m^{2}}^{(k-1) - 1} m^{(f) 2}(y) [\phi_{\alpha}^{(R)}(y)]^{2} - i e^{- l} \varphi^{(f)}(y) [\phi_{\alpha}^{(R)}(y)]^{2} \Big\} \nn && - \int d^{D} y Z_{\phi}^{(0)} \Big\{ \frac{N Z_{u}^{(0)}}{2 u^{(1)}(y)} [\varphi^{(0)}(y)]^{2} + \gamma_{2}^{(0)}(y) [m^{(0) 2}(y) - e^{- 2 l} m_{B}^{2}] + \gamma_{4}^{(0)}(y) [u^{(0)}(y) - e^{- (4 - D) l} u_{B}] \Big\} \nn && - \int d^{D} y \sum_{k = 1}^{f} Z_{\phi}^{(k-1)} \Big\{ l \frac{N \Pi_{p = 1}^{k} Z_{u}^{(p-1)}}{2 u^{(k)}(y)} \Big( \frac{\varphi^{(k)}(y) - \varphi^{(k-1)}(y)}{l} \Big)^{2} + l \frac{N \mathcal{C}_{\varphi}(m^{(k) 2})}{2} [\partial_{\mu} \varphi^{(k)}(y)]^{2} \nn && + \gamma_{2}^{(k)}(y) [m^{(k) 2}(y) - Z_{m^2}^{(k-1)} m^{(k-1) 2}(y)] + \gamma_{4}^{(k)}(y) [u^{(k)}(y) - Z_{\phi}^{(k-1)} Z_{u}^{(k-1)} u^{(k-1)}(y)] \Big\} \nn && + 2 \int d^{D} y \ln Z_{\phi}^{(0)} + 2 \int d^{D} y \sum_{k = 1}^{f} \ln Z_{\phi}^{(k-1)} \Big] . \eqa
We point out that $l \frac{N \mathcal{C}_{\varphi}(m^{(k) 2})}{2} [\partial_{\mu} \varphi^{(k)}(y)]^{2}$ is newly generated by quantum fluctuations of high-energy $\phi_{\alpha}^{(R)}(y)$ modes, given by the gradient expansion of their vacuum fluctuation energy with respect to the heavy mass at the $k-$th step of the RG transformation. $\mathcal{C}_{\varphi}(m^{(k) 2})$ is a positive parameter, which decreases as the mass increases.

%
%

Finally, we obtain an effective field theory with an emergent extra dimensional space, given by
\bqa && Z = \int D \phi_{\alpha}^{(R)}(y) D \varphi(y,z) D m^{2}(y,z) D \gamma_{2}(y,z) D u(y,z) D \gamma_{4}(y,z) \nn && \exp\Big[ - \int d^{D} y \frac{Z_{\phi}(y,z_{f})}{Z_{\phi}(y,0)} \Big\{ [\partial_{\mu} \phi_{\alpha}^{(R)}(y)]^{2} + \frac{Z_{m^{2}}^{-1}(y,z_{f})}{Z_{m^{2}}^{-1}(y,0)} m^{2}(y,z_{f}) [\phi_{\alpha}^{(R)}(y)]^{2} - i \varphi(y,z_{f}) [\phi_{\alpha}^{(R)}(y)]^{2} \Big\} \nn && - N \int d^{D} y Z_{\phi}(y,0) \Big\{ \frac{Z_{u}(y,0)}{2 u(y,0)} [\varphi(y,0)]^{2} - \gamma_{2}(y,0) \Big( \frac{\partial m^{2}(y,z)}{\partial z} \Big|_{z = 0} - 2 m^{2}(y,0) \Big) \nn && - \gamma_{4}(y,0) \Big( \frac{\partial u(y,z)}{\partial z} \Big|_{z = 0} - (4 - D) u(y,0) \Big) \Big\} - N \int_{0}^{z_{f}} d z \int d^{D} y Z_{\phi}(y,z) \Big\{ \frac{1}{2 u(y,z)} \frac{Z_{u}(y,z)}{Z_{u}(y,0)} [\partial_{z} \varphi(y,z)]^{2} \nn && + \frac{\mathcal{C}_{\varphi}[m^{2}(y,z)]}{2} [\partial_{\mu} \varphi(y,z)]^{2} - \gamma_{2}(y,z) \Big( \frac{\partial m^{2}(y,z)}{\partial z} - \frac{\partial \ln Z_{m^2}(y,z)}{\partial z} m^{2}(y,z) \Big) \nn && - \gamma_{4}(y,z) \Big( \frac{\partial u(y,z)}{\partial z} - \Big( \frac{\partial \ln Z_{\phi}(y,z)}{\partial z} + \frac{\partial \ln Z_{u}(y,z)}{\partial z} \Big) u(y,z) \Big) \Big\} + 2 \int d^{D} y \ln Z_{\phi}(y,z_{f}) \Big] , \label{RG_based_Dual_Holography} \eqa
where $N$ has been introduced in front of Lagrange multiplier fields in order to clarify the large $N$ limit later. It is clear the physical picture that this effective field theory describes. First of all, we point out that quantum fluctuations of the holographic or bulk dual order parameter field $\varphi(y,z)$ are suppressed in the large $N-$limit. This dual scalar field evolves through the emergent curved spacetime, given by the renormalized interaction vertex $u(y,z)$ and the renormalized mass $m^{2}(y,z)$ in $\mathcal{C}_{\varphi}[m^{2}(y,z)]$. We would like to emphasize that all the RG coefficients for fields and interaction vertices now depend on the order parameter field $\varphi(y,z)$, resulting in feedback effects to itself through the effective curved spacetime. In other words, both interaction and mass parameters are given by a function of the order parameter field. This intertwined renormalization gives rise to non-perturbative self-energy corrections in the dynamics of the original scalar field $\phi_{\alpha}^{(R)}(y)$ at IR $z = z_{f}$, described by $- i \varphi(y,z_{f}) [\phi_{\alpha}^{(R)}(y)]^{2}$. This is a novel large$-N$ mean-field theory with non-perturbative renormalization effects, which will be more clarified below. In the next section, we express all the RG coefficients in terms of an emergent metric tensor of an effective holographic dual field theory.

%
%

\section{Geometric encoding of renormalization group $\beta-$functions in an emergent dual holographic description for interacting relativistic bosons}

\subsection{Emergent dual holographic description}

Recently, we constructed or, more precisely, derived a dual holographic Einstein-Klein-Gordon-type effective field theory from an effective scalar field theory based on the recursive RG technique \cite{Holographic_Description_Einstein_Klein_Gordon} conceptually discussed in the previous section. We considered the following partition function
\bqa && Z = \int D \phi_{\alpha} \exp\Big\{ - \int d^{D} x \sqrt{g_{B}} \Big\{ g_{B}^{\mu\nu} (\partial_{\mu} \phi_{\alpha}) (\partial_{\nu} \phi_{\alpha}) + m^{2} \phi_{\alpha}^{2} + \xi R_{B} \phi_{\alpha}^{2} + \frac{u}{2N} \phi_{\alpha}^{2} \phi_{\beta}^{2} + \frac{\lambda}{2N} T_{\mu\nu} T^{\mu\nu} \Big\} \Big\} . \eqa
$\phi_{\alpha}$ is a scalar field with a flavor index $\alpha = 1, ..., N$. $g_{B}^{\mu\nu}$ is a background metric tensor, which may be $g_{B}^{\mu\nu} = \delta^{\mu\nu}$ to describe a $D-$dimensional Euclidean spacetime. $g_{B}$ is the determinant of the metric tensor, where $d^{D} x \sqrt{g_{B}}$ gives an invariant infinitesimal volume factor. $m$ is the mass of the scalar field, and $u$ is the strength of self-interactions. $R_{B}$ is Ricci scalar, where $\xi$ is the coupling constant between the scalar field and the scalar curvature \cite{Coupling_Scalarfields_Riccicurvature}. $T_{\mu\nu}$ is an energy-momentum tensor, where $\lambda$ is the coupling constant. Although this effective interaction is irrelevant at the Gaussian fixed point of this effective field theory, consideration of this tensor interaction plays an important role in the dual holographic description. It turns out that this effective interaction leads the metric tensor to be fully dynamical, to be discussed in more details.

To derive an effective dual holographic description for this effective field theory, we introduced two kinds of dual order parameter fields, which decompose both self-interactions of the density and the energy-momentum tensor, respectively. These dual fields are given by $\varphi(x,0)$ and $g^{\mu\nu}(x,0)$, respectively, where $0$ means the UV boundary $z = 0$ of the emergent extra dimensional space, corresponding to the absence of RG transformations. Then, we separate all quantum fields of $\phi_{\alpha}(x)$, $\varphi(x,0)$, and $g^{\mu\nu}(x,0)$ into their slow and fast degrees of freedom and perform the path integral for the fast degrees of freedom. As a result, we obtain an effective field theory for low-energy quantum fields, where various types of effective interactions are newly generated. For example, the Gaussian integration for both high-energy modes in $\varphi(x,0)$ and $g^{\mu\nu}(x,0)$ gives rise to additional self-interactions for both density and energy-momentum tensor channels. Once again, we perform the Hubbard-Stratonovich transformation for these newly generated self-interactions and obtain additional dual dynamical order parameter fields, given by $\varphi(x,1)$ and $g^{\mu\nu}(x,1)$. Here, $1$ denotes the first RG transformation. Considering additional effective interactions given by the integration of $\phi_{\alpha}(x)$, we can find a repetition rule for the recursive implementation of RG transformations.

Physical ingredients of this effective action after the first RG transformation are as follows. It is easy to figure out that quantum fluctuations of $\varphi(x,0)$ and $\varphi(x,1)$ are suppressed in the large $N$ limit while the path integral for original matter fields $\phi_{\alpha}(x)$ is Gaussian due to the second Hubbard-Stratonovich transformation and metric-tensor fields are also renormalized and updated. As a result, we obtain coupled mean-field equations for $\varphi(x,0)$ and $\varphi(x,1)$ with background geometry given by the upgraded metric tensor in the large $N$ limit. The mean-field equation of $\varphi(x,0)$ corresponds to the conventional saddle-point approximation of the related quantum field theory in the large $N$ limit. On the other hand, the mean-field equation of $\varphi(x,1)$ coupled to that of $\varphi(x,0)$ gives rise to $1/N$ corrections to the mean-field solution of $\varphi(x,0)$. This point has been demonstrated in Ref. \cite{Holographic_Description_Kondo_Effect} more explicitly. In addition, the renormalized metric tensor take into account vertex corrections in the one-loop level, where all coupling functions are renormalized. In this respect the effective action of the large $N$ limit after the first RG transformation is an effective mean-field theory for $\varphi(x,0)$, where both $1/N$ quantum corrections (self-energies) given by $\varphi(x,1)$ and vertex corrections described by the renormalized metric tensor have been introduced self-consistently through the RG transformation with background dual fields. In other words, this theoretical framework gives a one-loop-level RG-improved self-consistent mean-field theory in the large $N$ limit after the first RG transformation.

Repeating these RG transformations in a recursive way and taking the large $N$ limit, we find an effective mean-field theory for all collective dual order parameter fields, where not only $1/N$, $1/N^{2}$, ..., $1/N^{f}$ quantum corrections (self-energies) but also vertex corrections described by the renormalized metric tensor have been introduced and resumed self-consistently through these recursive RG transformations with background dual fields. Here, $f$ denotes the number of recursive RG transformations. It is important to notice that both self-energy and vertex corrections are self-consistently introduced and resumed to show an intertwined renormalization structure, which results in an effective large $N$ non-perturbative mean-field theory. In particular, this intertwined renormalization structure can be seen in the RG flow of the metric tensor, where the Green's function of high-energy quantum fluctuations of original matter fields takes into account the dynamical information, which depends on both the auxiliary dual field and the metric tensor field at the same time. This non-perturbative nature has been discussed in our recent study \cite{Intertwined_RG_Structure_KSK} more explicitly. We claim that this non-perturbative re-summation of the large $N$ limit is certainly a novel aspect of the present dual holographic description beyond all existing literatures based on the first-principle derivation.

The resulting holographic dual effective field theory is given by  \cite{Holographic_Description_Einstein_Klein_Gordon,Holographic_Description_Einstein_Maxwell}
\bqa && Z = \int D \varphi(x,z) D g_{\mu\nu}(x,z) \exp\Big[ - \mathcal{S}_{IR}[\varphi(x,z_{f}), g_{\mu\nu}(x,z_{f})] - \mathcal{S}_{Bulk}[\varphi(x,z), g_{\mu\nu}(x,z)] - \mathcal{S}_{UV}[\varphi(x,0), g_{\mu\nu}(x,0)] \Big] . \nn \label{Dual_Holographic_Theory_Partition_Function} \eqa
%
%
$\mathcal{S}_{IR}[\varphi(x,z_{f}), g_{\mu\nu}(x,z_{f})]$ is an effective action to give the IR boundary condition with the Gibbons-Hawking-York action \cite{Gibbons_Hawking_York_I,Gibbons_Hawking_York_II} from the bulk effective action. It is given by
\bqa && \mathcal{S}_{IR}[\varphi(x,z_{f}), g_{\mu\nu}(x,z_{f})] = - \ln \int D \phi_{\alpha}(x) \exp\Big[ - \int d^{D} x \sqrt{g(x,z_{f})} \Big\{ g^{\mu\nu}(x,z_{f}) [\partial_{\mu} \phi_{\alpha}(x)] [\partial_{\nu} \phi_{\alpha}(x)] \nn && + [m^{2} + \varphi(x,z_{f})] \phi_{\alpha}^{2}(x) \Big\} \Big] = N \int d^{D} x \sqrt{g(x,z_{f})} \Big\{ - \frac{\mathcal{C}_{\varphi}^{f}}{2} g^{\mu\nu}(x,z_{f}) [\partial_{\mu} \varphi(x,z_{f})] [\partial_{\nu} \varphi(x,z_{f})] - \mathcal{C}_{\xi}^{f} R(x,z_{f}) [\varphi(x,z_{f})]^{2} \nn && + \frac{1}{2 \kappa_{f}} \Big( R(x,z_{f}) - 2 \Lambda_{f} \Big) \Big\} , \label{Dual_Holographic_Theory_IR_Action} \eqa
where the gradient expansion with respect to $m$ has been performed. $\mathcal{C}_{\varphi}^{f}$ and $\mathcal{C}_{\xi}^{f}$ are positive numerical coefficients, which decrease as $m$ increases. The last Einstein-Hilbert action results from vacuum fluctuations of quantum matter fields, referred to as induced gravity \cite{Gradient_Expansion_Gravity_I,Gradient_Expansion_Gravity_II}. The holographic or bulk effective action is given by
\bqa && \mathcal{S}_{Bulk}[\varphi(x,z), g_{\mu\nu}(x,z)] = N \int_{0}^{z_{f}} d z \int d^{D} x \sqrt{g(x,z)} \Big\{ - \frac{1}{2u} [\partial_{z} \varphi(x,z)]^{2} - \frac{\mathcal{C}_{\varphi}}{2} g^{\mu\nu}(x,z) [\partial_{\mu} \varphi(x,z)] [\partial_{\nu} \varphi(x,z)] \nn && - \mathcal{C}_{\xi} R(x,z) [\varphi(x,z)]^{2} - \frac{1}{2 \lambda} \Big(\partial_{z} g^{\mu\nu}(x,z) - g^{\mu\alpha}(x,z) \big(\partial_{\alpha} \partial_{\alpha'} G_{xx'}[g_{\mu\nu}(x,z),\varphi(x,z)]\big)_{x' \rightarrow x} g^{\alpha'\nu}(x,z) \Big)^{2} \nn && + \frac{1}{2 \kappa} \Big( R(x,z) - 2 \Lambda \Big) \Big\} . \label{Dual_Holographic_Theory_Bulk_Action} \eqa
We have also seen a similar term of $- \frac{\sqrt{g(x,z)}}{2u} [\partial_{z} \varphi(x,z)]^{2}$ in the previous section, which results from the self-interaction of the density channel. $N \int_{0}^{z_{f}} d z \int d^{D} x \sqrt{g(x,z)} \Big\{ - \frac{\mathcal{C}_{\varphi}}{2} g^{\mu\nu}(x,z) [\partial_{\mu} \varphi(x,z)] [\partial_{\nu} \varphi(x,z)] - \mathcal{C}_{\xi} R(x,z) [\varphi(x,z)]^{2} + \frac{1}{2 \kappa} \Big( R(x,z) - 2 \Lambda \Big) \Big\}$ comes from vacuum fluctuations of high-energy modes in $\phi_{\alpha}(x)$ \cite{Gradient_Expansion_Gravity_I,Gradient_Expansion_Gravity_II} during the recursive RG transformation, as discussed in the IR effective action. The most nontrivial term is $- \frac{1}{2 \lambda} \Big(\partial_{z} g^{\mu\nu}(x,z) - g^{\mu\alpha}(x,z) \big(\partial_{\alpha} \partial_{\alpha'} G_{xx'}[g_{\mu\nu}(x,z),\varphi(x,z)]\big)_{x' \rightarrow x} g^{\alpha'\nu}(x,z) \Big)^{2}$ to describe the evolution of the metric tensor in the extra dimensional space. Here, $G_{xx'}[g_{\mu\nu}(x,z),\varphi(x,z)]$ is the Green's function of heavy scalar fields, given by
\bqa && \Big\{- \frac{1}{\sqrt{g(x,z)}} \partial_{\mu} \Big( \sqrt{g(x,z)} g^{\mu\nu}(x,z) \partial_{\nu} \Big) + \frac{m^{2} + \varphi(x,z)}{\epsilon} \Big\} G_{xx'}[g_{\mu\nu}(x,z),\varphi(x,z)] = \frac{1}{\sqrt{g(x,z)}} \delta^{(D)}(x-x') , \nn \label{Dual_Holographic_Theory_Green_Function} \eqa
where the infinitesimal parameter $\epsilon \sim d l$ denotes the heavy mass character. We emphasize that only this Green's function encodes the information on dynamics. On the other hand, the second-order $z-$derivative in the metric tensor results from the self-interaction of the energy-momentum tensor channel. Below, we consider the case of $\lambda \rightarrow 0$, which gives rise to $\partial_{z} g^{\mu\nu}(x,z) = g^{\mu\alpha}(x,z) \big(\partial_{\alpha} \partial_{\alpha'} G_{xx'}[g_{\mu\nu}(x,z),\varphi(x,z)]\big)_{x' \rightarrow x} g^{\alpha'\nu}(x,z)$. We point out our gauge fixing for the metric tensor \cite{Holographic_Description_Einstein_Klein_Gordon,Holographic_Description_Einstein_Maxwell}, given by $g^{DD}(x,z) = 0$ and $g^{\mu D}(x,z) = g^{D \nu}(x,z) = 1$ with $\mu, ~ \nu = 0, ..., D - 1$ and involved with the lapse function and the shift vector in the Hamiltonian formulation of the metric tensor, respectively \cite{ADM_Formulation}. The UV effective action is
\bqa && \mathcal{S}_{UV}[\varphi(x,0), g_{\mu\nu}(x,0)] = N \int d^{D} x \sqrt{g(x,0)} \Big\{ - \frac{1}{2u} \Big( \varphi(x,0) - \xi R(x,0) \Big)^{2} - \frac{1}{2 \lambda} \Big(g^{\mu\nu}(x,0) - g_{B}^{\mu\nu}(x)\Big)^{2} \Big\} , \label{Dual_Holographic_Theory_UV_Action} \eqa
which determines the UV boundary condition with the Gibbons-Hawking-York action from the above bulk effective action. One may have $g_{B}^{\mu\nu}(x) = \delta^{\mu\nu}$ and $R(x,0) = 0$ as a UV boundary condition for the metric tensor.

To compare this emergent dual holographic description with the RG-reformulated effective field theory of Eq. (\ref{RG_based_Dual_Holography}), we focus on the limit of $\lambda \rightarrow 0$. Then, we obtain the following dual holographic effective field theory
\bqa && Z = \int D \phi_{\alpha}(x) D \varphi(x,z) D g_{\mu\nu}(x,z) \delta\Big(g^{\mu\nu}(x,0) - g_{B}^{\mu\nu}(x)\Big) \nn && \delta\Big\{\partial_{z} g^{\mu\nu}(x,z) - g^{\mu\alpha}(x,z) \Big(\partial_{\alpha} \partial_{\alpha'} G_{xx'}[g_{\mu\nu}(x,z),\varphi(x,z)]\Big)_{x' \rightarrow x} g^{\alpha'\nu}(x,z) \Big\} \nn && \exp\Big[ - \int d^{D} x \sqrt{g(x,z_{f})} \Big\{ g^{\mu\nu}(x,z_{f}) [\partial_{\mu} \phi_{\alpha}(x)] [\partial_{\nu} \phi_{\alpha}(x)] + [m^{2} + \varphi(x,z_{f})] \phi_{\alpha}^{2}(x) \Big\} \nn && - N \int d^{D} x \sqrt{g(x,0)} \Big\{- \frac{1}{2u} [\varphi(x,0)]^{2} \Big\} \nn && - N \int_{0}^{z_{f}} d z \int d^{D} x \sqrt{g(x,z)} \Big\{- \frac{1}{2u} [\partial_{z} \varphi(x,z)]^{2} - \frac{\mathcal{C}_{\varphi}}{2} g^{\mu\nu}(x,z) [\partial_{\mu} \varphi(x,z)] [\partial_{\nu} \varphi(x,z)] + \frac{1}{2 \kappa} \Big( R(x,z) - 2 \Lambda \Big) \Big\} \Big] . \nn \label{GR_Holography} \eqa
An essential point is that the RG flow of the metric tensor through the extra dimensional space is now the first order in the $z-$derivative. In other words, the metric tensor is not fully dynamical any more. We also point out that this mathematical form of the RG flow for the metric tensor remains to be almost unchanged in the absence of any interactions, given by the limit of $u \rightarrow 0$ and reduced to be
\bqa && Z = \int D g_{\mu\nu}(x,z) \delta\Big(g^{\mu\nu}(x,0) - g_{B}^{\mu\nu}(x)\Big) \delta\Big\{ \partial_{z} g^{\mu\nu}(x,z) - g^{\mu\alpha}(x,z) \Big(\partial_{\alpha} \partial_{\alpha'} G_{xx'}[g_{\mu\nu}(x,z)]\Big)_{x' \rightarrow x} g^{\alpha'\nu}(x,z) \Big\} \nn && \exp\Big[ - \frac{N}{2 \kappa_{f}}  \int d^{D} x \sqrt{g(x,z_{f})} \Big( R(x,z_{f}) - 2 \Lambda_{f} \Big) - \frac{N}{2 \kappa} \int_{0}^{z_{f}} d z \int d^{D} x \sqrt{g(x,z)} \Big( R(x,z) - 2 \Lambda \Big) \Big] , \eqa
where the path integral for matter fields has been taken and the Green's function of high-energy boson fields is
\bqa && \Big\{- \frac{1}{\sqrt{g(x,z)}} \partial_{\mu} \Big( \sqrt{g(x,z)} g^{\mu\nu}(x,z) \partial_{\nu} \Big) + \frac{m^{2}}{\epsilon} \Big\} G_{xx'}[g_{\mu\nu}(x,z),\varphi(x,z)] = \frac{1}{\sqrt{g(x,z)}} \delta^{(D)}(x-x') . \nonumber \eqa
We emphasize that this geometric description is just a reformulation of a free boson theory, where there exist no dynamical degrees of freedom in the metric tensor. It turns out that the evolution equation of the metric tensor describes the RG flow of the hopping parameter in a non-interacting lattice model \cite{Holographic_Description_Einstein_Klein_Gordon,Intertwined_RG_Structure_KSK}. In this respect an important point for the description of non-perturbative physics is the intertwined renormalization structure between the order parameter field and the interaction vertices or the background geometry \cite{Intertwined_RG_Structure_KSK}. Furthermore, effective interactions between energy-momentum tensor currents uplift the metric tensor to be fully dynamical, encoded into the second-order $z-$derivative in the bulk effective action.

%
%

\subsection{How to identify renormalization group $\beta-$functions with background gravity}

Finally, we express all the RG coefficients of the RG-reformulated effective field theory in terms of the metric tensor of the emergent dual holographic theory. We recall the RG-reformulated effective field theory of Eq. (\ref{RG_based_Dual_Holography}) with $Z_{\phi}(y,0) = Z_{m^{2}}(y,0) = Z_{u}(y,0) = 1$,
\bqa && Z = \int D \phi_{\alpha}^{(R)}(y) D \varphi(y,z) \exp\Big[ - \int d^{D} y Z_{\phi}(y,z_{f}) \Big\{ [\partial_{\mu} \phi_{\alpha}^{(R)}(y)]^{2} + [Z_{m^{2}}^{-1}(y,z_{f}) m^{2}(y,z_{f}) + \varphi(y,z_{f})] [\phi_{\alpha}^{(R)}(y)]^{2} \Big\} \nn && - N \int d^{D} y \Big\{- \frac{1}{2 u} [\varphi(y,0)]^{2} \Big\} - N \int_{0}^{z_{f}} d z \int d^{D} y Z_{\phi}(y,z) \Big\{- \frac{Z_{u}(y,z)}{2 u(y,z)} [\partial_{z} \varphi(y,z)]^{2} - \frac{\mathcal{C}_{\varphi}[m^{2}(y,z)]}{2} [\partial_{\mu} \varphi(y,z)]^{2} \Big\} \Big] , \nonumber \eqa
where the Lagrange multiplier fields were integrated out to give the constraint equations of RG $\beta-$functions as
\bqa && \frac{\partial \ln m^{2}(y,z)}{\partial z} = \frac{\partial \ln Z_{m^2}(y,z)}{\partial z} , ~~~~~ \frac{\partial \ln u(y,z)}{\partial z} = \frac{\partial \ln Z_{\phi}(y,z)}{\partial z} + \frac{\partial \ln Z_{u}(y,z)}{\partial z} \label{RG_Beta_Functions_Formal} \eqa
with the initial conditions of $\frac{\partial \ln m^{2}(y,z)}{\partial z} \Big|_{z = 0} = 2$ and $\frac{\partial \ln u(y,z)}{\partial z} \Big|_{z = 0} = (4 - D)$.

To compare this RG-reformulated effective field theory with the holographic dual effective field theory, we consider an effective renormalized action. Here, we separate both the mass term of the IR effective action $Z_{m^{2}}^{-1}(y,z_{f}) m^{2}(y,z_{f}) [\phi_{\alpha}^{(R)}(y)]^{2}$ and the RG-flow term of the bulk action $- \frac{Z_{u}(y,z)}{2 u(y,z)} [\partial_{z} \varphi(y,z)]^{2}$ into their renormalized and counter-term parts as follows $Z_{m^{2}}^{-1}(y,z_{f}) m^{2}(y,z_{f}) [\phi_{\alpha}^{(R)}(y)]^{2} = m^{2}(y,z_{f}) [\phi_{\alpha}^{(R)}(y)]^{2} + [Z_{m^{2}}^{-1}(y,z_{f}) - 1] m^{2}(y,z_{f}) [\phi_{\alpha}^{(R)}(y)]^{2}$ and $- \frac{Z_{u}(y,z)}{2 u(y,z)} [\partial_{z} \varphi(y,z)]^{2} = - \frac{1}{2 u(y,z)} [\partial_{z} \varphi(y,z)]^{2} - \frac{Z_{u}(y,z) - 1}{2 u(y,z)} [\partial_{z} \varphi(y,z)]^{2}$, and keep only the first terms in the renormalized effective action.
%
%
%
Resorting to the formal RG $\beta-$functions Eq. (\ref{RG_Beta_Functions_Formal}), we consider the following RG-reformulated renormalized field theory
\bqa && Z_{R} = \int D \phi_{\alpha}^{(R)}(y) D \varphi(y,z) \exp\Big[ - \int d^{D} y Z_{\phi}(y,z_{f}) \Big\{ [\partial_{\mu} \phi_{\alpha}^{(R)}(y)]^{2} + [Z_{m^{2}}(y,z_{f}) m^{2} + \varphi(y,z_{f})] [\phi_{\alpha}^{(R)}(y)]^{2} \Big\} \nn && - N \int d^{D} y \Big\{- \frac{1}{2 u} [\varphi(y,0)]^{2} \Big\} - N \int_{0}^{z_{f}} d z \int d^{D} y Z_{\phi}(y,z) \Big\{- \frac{Z_{u}(y,z)}{2 u} [\partial_{z} \varphi(y,z)]^{2} - \frac{\mathcal{C}_{\varphi}[m^{2}(y,z)]}{2} [\partial_{\mu} \varphi(y,z)]^{2} \Big\} \Big] , \nn \label{RG_Holography_Ansatz} \eqa
where both the renormalized mass and the interaction parameter have been replaced with the RG coefficients of $Z_{m^{2}}(y,z_{f})$ and $Z_{u}(y,z)$, respectively.

Taking the rotational symmetric ansatz for the metric tensor
\bqa && g^{\mu\nu}(x,z) = f(x,z) \delta^{\mu\nu} , \eqa
Eq. (\ref{GR_Holography}) leads to an effective holographic dual field theory
\bqa && Z = \int D \phi_{\alpha}(x) D \varphi(x,z) \exp\Big[ - \int d^{D} x f^{-D/2}(x,z_{f}) \Big\{ f(x,z_{f}) [\partial_{\mu} \phi_{\alpha}(x)]^{2} + [m^{2} + \varphi(x,z_{f})] \phi_{\alpha}^{2}(x) \Big\} \nn && - N \int d^{D} x \Big\{- \frac{1}{2u} [\varphi(x,0)]^{2} \Big\} - N \int_{0}^{z_{f}} d z \int d^{D} x f^{-D/2}(x,z) \Big\{- \frac{1}{2u} [\partial_{z} \varphi(x,z)]^{2} - \frac{\mathcal{C}_{\varphi}}{2} f(x,z) [\partial_{\mu} \varphi(x,z)]^{2} \nn && + \frac{1}{2 \kappa} \Big( R(x,z) - 2 \Lambda \Big) \Big\} \Big] . \label{GR_Holography_Ansatz} \eqa
Here, the metric function $f(x,z)$ is governed by the RG-flow equation
\bqa && \partial_{z} f(x,z) = \Big( - \partial_{\mu}^{2} G_{xx'}[f(x,z),\varphi(x,z)] \Big)_{x' \rightarrow x} [f(x,z)]^{2} \eqa
with the initial condition $f(x,0) = 1$, where the Green's function is
\bqa && \Big\{f(x,z) (- \partial_{\mu}^{2}) - \Big( 1 - \frac{D}{2} \Big) \Big( \partial_{\mu} f(x,z) \Big) \partial_{\mu} + \frac{m^{2} + \varphi(x,z)}{\epsilon} \Big\} G_{xx'}[f(x,z),\varphi(x,z)] = f^{D/2}(x,z) \delta^{(D)}(x-x') . \nn \eqa

Comparing Eq. (\ref{RG_Holography_Ansatz}) with Eq. (\ref{GR_Holography_Ansatz}), finally we express all the RG coefficients with the metric function $f(x,z_{f})$ and the order-parameter field $\varphi(x,z_{f})$, given by
\bqa && Z_{\phi}(x,z_{f}) = f^{-\frac{D}{2}+1}(x,z_{f}) \eqa
for the field renormalization,
\bqa && Z_{m^{2}}(x,z_{f}) = \Big( 1 + \frac{\varphi(x,z_{f})}{m^{2}} \Big) f^{-D/2}(x,z_{f}) - \frac{\varphi(x,z_{f})}{m^{2}} \eqa
for the mass renormalization, and
\bqa && Z_{u}(x,z) = f^{- 1}(x,z) \eqa
for the interaction renormalization. These equations result in the RG $\beta-$functions
\bqa && \beta_{m^{2}} \equiv \frac{\partial \ln m^{2}(x,z_{f})}{\partial z_{f}} = \frac{\partial }{\partial z_{f}} \ln \Big\{ \Big( 1 + \frac{\varphi(x,z_{f})}{m^{2}} \Big) f^{-D/2}(x,z_{f}) - \frac{\varphi(x,z_{f})}{m^{2}} \Big\} , \nn && \beta_{u} \equiv \frac{\partial \ln u(x,z_{f})}{\partial z_{f}} = - \frac{D}{2} \frac{\partial \ln f(x,z_{f})}{\partial z_{f}} . \eqa
We also have $\mathcal{C}_{\varphi}[m^{2}(x,z)] = \mathcal{C}_{\varphi}$. Here, the equation of motion for the order-parameter field is given by the Euler-Lagrange equation of motion for $\varphi(x,z)$ from Eq. (\ref{GR_Holography_Ansatz}),
\bqa && 0 = \frac{1}{u} \Big(- \partial_{z}^{2} \varphi(x,z) + \frac{D}{2} [\partial_{z} \ln f(x,z)] \partial_{z} \varphi(x,z) \Big) + \mathcal{C}_{\varphi} f(x,z) \Big\{ - \partial_{\mu}^{2} \varphi(x,z) + \Big( \frac{D}{2}-1 \Big) [\partial_{\mu} \ln f(x,z)] \partial_{\mu} \varphi(x,z) \Big\} . \nn \eqa
The UV boundary condition gives $\varphi(x,0) = 0$ while the IR boundary condition leads to
\bqa && \frac{D}{4u} [\partial_{z} \ln f(x,z)]_{z = z_{f}} \varphi(x,z_{f}) = \Big(G_{xx'}^{\phi}[f(x,z_{f}),\varphi(x,z_{f})]\Big)_{x' \rightarrow x} . \eqa
This IR boundary condition should be regarded as an effective mean-field equation in the large $N$ limit, where all the coefficients of mass and interactions are renormalized and described by the renormalized metric function $f(x,z_{f})$. The IR Green's function is
\bqa && \Big\{ - \partial_{\mu}^{2} + \Big( \frac{D}{2}-1 \Big) [\partial_{\mu} \ln f(x,z_{f})] \partial_{\mu} + \frac{m^{2} + \varphi(x,z_{f})}{f(x,z_{f})} \Big\} G_{xx'}^{\phi}[f(x,z_{f}),\varphi(x,z_{f})] = f^{\frac{D}{2}-1}(x,z_{f}) \delta^{(D)}(x-x') , \nn \eqa
which determines the renormalized value of the order-parameter field $\varphi(x,z_{f})$. This completes the proof of equivalence between the two partition functions of Eqs. (\ref{RG_Holography_Ansatz}) and (\ref{GR_Holography_Ansatz}).

\section{Discussion: Ricci flow}

Our emergent dual holographic theory is given by the partition function of Eq. (\ref{Dual_Holographic_Theory_Partition_Function}), which consists of the holographic effective action Eq. (\ref{Dual_Holographic_Theory_Bulk_Action}), the IR effective action Eq. (\ref{Dual_Holographic_Theory_IR_Action}) with the UV effective action Eq. (\ref{Dual_Holographic_Theory_UV_Action}), and the kernel Eq. (\ref{Dual_Holographic_Theory_Green_Function}) for the evolution of the metric tensor. Here, only the Green's function Eq. (\ref{Dual_Holographic_Theory_Green_Function}) encodes the dynamics information of quantum matter fields while all other ingredients are geometrized. One may try to geometrize even this dynamics information, reformulating the Green's function in terms of the metric and curvature tensors. Recalling the evolution equation for the metric tensor in the absence of interactions between energy-momentum tensor currents, an interesting observation is that the second-order derivative of the Green's function with respect to the $D-$dimensional spacetime appears. Intuitively, it is natural trying to identify this quantity with an effective curvature. In the context of the fractional quantum Hall effect, this identification has been pointed out \cite{FQHE_Geometry}. This discussion reminds us of Ricci flow.

The Ricci flow equation \cite{Ricci_Flow_0,Ricci_Flow_I,Ricci_Flow_II,Ricci_Flow_III,Ricci_Flow_IV,Ricci_Flow_V} is to describe the deformation of a Riemannian metric $g_{\mu\nu}(x,z)$ with an extra-dimensional space coordinate $z$, here, which plays the same role as time. $\mu$ and $\nu$ cover the $D-$dimensional spacetime coordinate $0, ..., D-1$. This evolution equation may be regarded as an analog of the diffusion equation for geometries, given by a parabolic partial differential equation. The deformation is governed by Ricci curvature, which leads to homogeneity of geometry. We conjecture that the evolution equation for the metric tensor in the emergent dual holographic description is given by the Ricci flow equation
\bqa && \partial_{z} g_{\mu\nu}(x,z) = - 2 R_{\mu\nu}(x,z) , \eqa
%
%
where $R_{\mu\nu}(x,z)$ is the Ricci curvature tensor. As a result, the dual holographic theory is given by the following partition function
\bqa && Z = Z_{\Lambda} \int D \varphi(x,z) D g_{\mu\nu}(x,z) \nn && \exp\Big[ - N \int_{0}^{z_{f}} d z \int d^{D} x \sqrt{g(x,z)} \Big\{ - \frac{1}{2u} [\partial_{z} \varphi(x,z)]^{2} - \frac{\mathcal{C}_{\varphi}}{2} g^{\mu\nu}(x,z) [\partial_{\mu} \varphi(x,z)] [\partial_{\nu} \varphi(x,z)] - \mathcal{C}_{\xi} R(x,z) [\varphi(x,z)]^{2} \nn && - \frac{1}{2 \lambda} \Big(\partial_{z} g^{\mu\nu}(x,z) + 2 R^{\mu\nu}(x,z) \Big) \Big(\partial_{z} g_{\mu\nu}(x,z) + 2 R_{\mu\nu}(x,z) \Big) + \frac{1}{2 \kappa} \Big( R(x,z) - 2 \Lambda \Big) \Big\} \nn && - N \int d^{D} x \sqrt{g(x,0)} \Big\{- \frac{1}{2u} \Big( \varphi(x,0) - \xi R(x,0) \Big)^{2} - \frac{1}{2 \lambda} \Big(g^{\mu\nu}(x,0) - g_{B}^{\mu\nu}(x)\Big)^{2} \Big\} \nn && - N \int d^{D} x \sqrt{g(x,z_{f})} \Big\{ - \frac{\mathcal{C}_{\varphi}^{f}}{2} g^{\mu\nu}(x,z_{f}) [\partial_{\mu} \varphi(x,z_{f})] [\partial_{\nu} \varphi(x,z_{f})] - \mathcal{C}_{\xi}^{f} R(x,z_{f}) [\varphi(x,z_{f})]^{2} \nn && + \frac{1}{2 \kappa_{f}} \Big( R(x,z_{f}) - 2 \Lambda_{f} \Big) \Big\} \Big] . \label{Ricci_Flow_Holographic_Theory} \eqa
%
%
Now, the quantum field theory is both holographically and geometrically dualized.

Here, we have to point out that the holographic RG flow gives rise to the Ricci flow naturally \cite{Holographic_RG_Flow_Ricci_Flow_I,Holographic_RG_Flow_Ricci_Flow_II}. Actually, the holographic dual effective field theory Eq. (\ref{Dual_Holographic_Theory_Bulk_Action}) turns out to reproduce this result at the fixed point of the $z_{f} \rightarrow \infty$ limit. To verify this statement, we consider the Hamilton-Jacobi formulation \cite{Holographic_Duality_IV,Holographic_Duality_V, Holographic_Duality_VI} for the holographic dual effective field theory Eq. (\ref{Dual_Holographic_Theory_Bulk_Action}), given by
\bqa && \frac{1}{\sqrt{g(x,z_{f})}} \frac{\partial}{\partial z_{f}} \mathcal{I}[\varphi(x,z_{f}), g_{\mu\nu}(x,z_{f})] = \frac{N u}{2} \Big\{ \frac{1}{\sqrt{g(x,z_{f})}} \frac{\partial \mathcal{I}[\varphi(x,z_{f}), g_{\mu\nu}(x,z_{f})]}{\partial \varphi(x,z_{f})} \Big\}^{2} \nn && - \frac{N \mathcal{C}_{\varphi}}{2} g^{\mu\nu}(x,z_{f}) [\partial_{\mu} \varphi(x,z_{f})] [\partial_{\nu} \varphi(x,z_{f})] - N \mathcal{C}_{\xi} R(x,z_{f}) [\varphi(x,z_{f})]^{2} + \frac{N \lambda}{2} \Big\{ \frac{1}{\sqrt{g(x,z_{f})}} \frac{\partial \mathcal{I}[\varphi(x,z_{f}), g_{\mu\nu}(x,z_{f})]}{\partial g^{\mu\nu}(x,z_{f})} \Big\}^{2} \nn && + N \beta_{\mu\nu}^{g}[\varphi(x,z_{f}), g_{\mu\nu}(x,z_{f})] \Big\{ \frac{1}{\sqrt{g(x,z_{f})}} \frac{\partial \mathcal{I}[\varphi(x,z_{f}), g_{\mu\nu}(x,z_{f})]}{\partial g^{\mu\nu}(x,z_{f})} \Big\} + \frac{N}{2 \kappa} \Big( R(x,z_{f}) - 2 \Lambda \Big) . \label{Hamilton_Jacobi_Eq} \eqa
Here, $\mathcal{I}[\varphi(x,z_{f}), g_{\mu\nu}(x,z_{f})]$ is an effective IR action and \bqa && \beta_{g}^{\mu\nu}[\varphi(x,z_{f}), g_{\mu\nu}(x,z_{f})] = g^{\mu\alpha}(x,z_{f}) \big(\partial_{\alpha} \partial_{\alpha'} G_{xx'}[g_{\mu\nu}(x,z_{f}),\varphi(x,z_{f})]\big)_{x' \rightarrow x} g^{\alpha'\nu}(x,z_{f}) \nonumber \eqa is the $\beta-$function for the RG flow of the metric tensor, which has to vanish in the $z_{f} \rightarrow \infty$ limit corresponding to a fixed point. When this $\beta-$function of the metric tensor vanishes at the fixed point, this Hamilton-Jacobi equation is essentially reduced into that of the holographic duality conjecture \cite{Holographic_Duality_IV,Holographic_Duality_V, Holographic_Duality_VI}, which reproduces the Ricci flow \cite{Holographic_RG_Flow_Ricci_Flow_I,Holographic_RG_Flow_Ricci_Flow_II}. On the other hand, when the $\beta_{g}^{\mu\nu}[\varphi(x,z_{f}), g_{\mu\nu}(x,z_{f})]$ function does not vanish, it depends on both $\varphi(x,z_{f})$ and $g_{\mu\nu}(x,z_{f})$ through the Green's function in an intertwined and nonlinear way, and this nonlinear intertwined renormalization structure gives rise to difficulty in solving this Hamilton-Jacobi equation. This is the reason why we reformulate the RG transformation in a recursive way in order to determine RG $\beta-$functions non-perturbatively from our holographic dual effective field theory. The present prescription allows us to find all the RG-transformation coefficients even away from quantum criticality (at any length scale), regarded to generalize the holographic RG flow of the AdS geometry \cite{Holographic_RG_Wilson_RG_Equivalence} towards that without conformal symmetry \cite{Holographic_Description_Entanglement_Entropy_Comments}.

\section{Conclusion}

The emergent holographic dual effective field theory Eq. (\ref{Dual_Holographic_Theory_Partition_Function}), given by the holographic bulk effective action Eq. (\ref{Dual_Holographic_Theory_Bulk_Action}), the IR effective action Eq. (\ref{Dual_Holographic_Theory_IR_Action}) with the UV effective action Eq. (\ref{Dual_Holographic_Theory_UV_Action}), and the kernel Eq. (\ref{Dual_Holographic_Theory_Green_Function}) for the evolution of the metric tensor, is an effective mean-field theory in the large $N$ limit, which turns out to resum quantum corrections in the all-loop order self-consistently. Here, the RG flow of the metric tensor describes renormalization effects of interaction parameters, regarded as vertex corrections and intertwined with the Euler-Lagrange equation of motion of the order-parameter field. This intertwined renormalization structure between vertex and self0energy corrections is realized in the IR boundary condition, which corresponds to an effective mean-field equation, where all interaction parameters are self-consistently renormalized and described by the emergent background curved spacetime metric, responsible for the description of non-perturbative physics in this dual holographic effective field theory.

Based on this dual holographic effective field theory, we proposed a prescription on how to find RG $\beta-$functions in a non-perturbative way. Although the holographic renormalization method has been utilized to determine such RG $\beta-$functions non-perturbatively in the holographic duality conjecture, it turns out that it is not easy to solve the Hamilton-Jacobi equation of the holographic renormalization framework in our holographic dual effective field theory. The origin of this difficulty lies at the self-consistently intertwined renormalization structure between the metric tensor and the order-parameter field away from quantum criticality, given by the $\beta-$function of the metric tensor in Eq. (\ref{Hamilton_Jacobi_Eq}) and not described by the AdS geometry. In this respect we reformulated the RG transformation in a recursive way and obtained the formal expression Eq. (\ref{RG_based_Dual_Holography}), which manifests the nonlinearly intertwined RG flow of the effective action through the emergent extra dimensional space. As a result, we could find all the RG coefficients in terms of the emergent metric tensors non-perturbatively and express the RG $\beta-$functions with the metric function even away from quantum criticality without conformal symmetry. This prescription generalizes the holographic RG flow of the holographic duality conjecture in the present holographic dual effective field theory. Furthermore, we conjectured that the RG $\beta-$functions may be related with the Ricci flow equation of the metric tensor. Replacing the evolution equation of the metric tensor with the Ricci flow equation, we proposed a both holographically and geometrically dualized effective field theory for a quantum field theory.

It has been proposed that the Ricci flow constrains the RG flow of the entanglement entropy for a black hole, resulting in monotonicity \cite{Ricci_Flow_Entanglement_Entropy}. Recently, we calculated the entanglement entropy based on this effective dual holographic theory and found entanglement transfer from quantum matter to classical geometry \cite{Holographic_Description_Einstein_Klein_Gordon}. We claimed that this entanglement transfer serves as a microscopic foundation to the emergent holographic description. It would be interesting to calculate the entanglement entropy of Eq. (\ref{Ricci_Flow_Holographic_Theory}), where the Ricci flow may constrain the entanglement transfer.

\section*{Acknowledgement}

K.-S. Kim was supported by the Ministry of Education, Science, and Technology (No. 2011-0030046) of the National Research Foundation of Korea (NRF) and by TJ Park Science Fellowship of the POSCO TJ Park Foundation. K.-S. Kim appreciates fruitful discussions with Shinsei Ryu and his hospitality during the sabbatical leave.

\end{document}